\documentclass[%
 aip,
 amsmath,amssymb,
preprint,%
]{revtex4-1}

\usepackage{graphicx}
\usepackage{dcolumn}
\usepackage{bm}

\usepackage[utf8]{inputenc}
\usepackage[T1]{fontenc}
\usepackage{mathptmx}
\usepackage{hyperref}
\usepackage{xcolor}
\usepackage{color,soul}

\begin{document}

\preprint{AIP/123-QED}

\title{Characterization of superconducting NbTiN films using a Dispersive Fourier Transform Spectrometer}

\author{B.N.R. Lap}
\email[Authors to whom correspondence should be addressed: ]{Bram Lap, b.lap@sron.nl; Andrey Khudchenko, khudchenko@asc.rssi.ru}
\affiliation{ SRON Netherlands Institute for Space Research, Groningen, The Netherlands}
\affiliation{ Kapteyn Astronomical Insitute, University of Groningen, Groningen, The Netherlands}

\author{A. Khudchenko}
\affiliation{ Astro Space Center of P.N. Lebedev Physical Institute RAS, Moscow, Russia}

\author{R. Hesper }
\affiliation{ Kapteyn Astronomical Insitute, University of Groningen, Groningen, The Netherlands}

\author{K. I. Rudakov}
\affiliation{ Kapteyn Astronomical Insitute, University of Groningen, Groningen, The Netherlands}
\affiliation{ SRON Netherlands Institute for Space Research, Groningen, The Netherlands}
\affiliation{ Kotel’nikov Institute of Radio Engineering and Electronics RAS, Moscow, Russia}

\author{P. Dmitriev}
\affiliation{ Kotel’nikov Institute of Radio Engineering and Electronics RAS, Moscow, Russia}

\author{F.Khan}
\affiliation{ Kotel’nikov Institute of Radio Engineering and Electronics RAS, Moscow, Russia}
\affiliation{ Moscow Institute of Physics and Technology, Dolgoprudny, Russia  }

\author{V. Koshelets} 
\affiliation{ Kotel’nikov Institute of Radio Engineering and Electronics RAS, Moscow, Russia}

\author{A. M. Baryshev}
\affiliation{ Kapteyn Astronomical Insitute, University of Groningen, Groningen, The Netherlands}

\date{\today}

\begin{abstract}
We have built a Terahertz Dispersive Fourier Transform Spectrometer \cite{Birch1987} to study frequency properties of superconducting films used for fabrication of THz detectors. The signal reflected from the tested film is measured in time domain, which allows to separate it from the other reflections. The complex conductivity of the film depends on frequency and determines the reflection coefficient. By comparing the film reflection in the superconducting state (temperature is below $T_c$) with the reflection of the normal state, we characterise the film quality at terahertz frequencies. The method was applied to 70 and 200nm thick Nb films on a silicon wafer and to 360nm thick NbTiN films on silicon and quartz wafers. The strong-coupling coefficient, $\alpha$, was found to be 3.52 for Nb, and 3.71-4.02 for the NbTiN films. The experimental results were fitted using extended Mattis-Bardeen theory \cite{Noguchi2012} and show a good agreement. 
\end{abstract}

\maketitle

Superconductor-insulator-superconductor (SIS) terahertz (THz) mixers are used in most submillimeter telescopes around the world, such as Atacama Large Millimeter Array \cite{ALMAlink} and many more \cite{gusten2006apex,chenu2016front,SMAlink,Greenland_link}, because they have the lowest available noise. In order to achieve quantum limited performance, the micron size SIS device must be embedded into an on-chip low-loss superconducting resonant matching circuit \cite{tucker1985quantum,d1984sis}, but the quality of the used materials strongly influences the performance of the device. In this paper, we propose and experimentally demonstrate a Dispersive Fourier Transform Spectrometer \cite{Birch1987} (DFTS) scheme used to characterise the performance of representative superconducting film structures in the equilibrium state at Terahertz frequencies. We aim to test the base electrode of the NbTiN film, which are used as microstrip circuits of the state-of-the-art SIS receivers for $\nu > 750$ GHz \cite{Uzawa2015,Jackson2006}. The film can be measured after each technological step, allowing us to investigate the impact of production process on the key film parameters. An RF current is used for the measurements and we only investigate the operating top layer of the film, i.e. the bottom electrode in the tuning circuit of our SIS junction.

Transmission Time Domain Spectroscopy (TDS) is the conventional technique for studying the properties of superconducting films \cite{Mittleman1996,Laamiri_2009}, but this technique has a major drawback: it requires thin films, i.e. a thickness, $d < \lambda_L$, where $\lambda_L$ id the London penetration depth \cite{Mittleman1996,Laamiri_2009,Uzawa2015}, but for actual SIS devices $d > \lambda_L$. This makes a comparison between the film measured by TDS and the film used in the SIS mixers more complex. An alternative is to use on reflection measurements, for which there are three main options: Backward-Wave Oscillators (BWOs) \cite{gorshunov2007,parshin2020}, TDS and DFTS. BWOs are limited in frequency (to $\nu \sim 1.5 \text{THz}$) and experimentally demanding, because it requires a set of local oscillators between one has to switch during the experiment. In general, there is no real difference between reflection TDS and (D)FTS \cite{Han2001,Savini2010,Savini2016} measurements. However, measuring changes in the order of $\sim 1$\%, which is required for characterizing SIS structures, in reflection is difficult for TDS due to stability issues \cite{Guo2019}, therefore we opted for DFTS. 

\begin{figure*}
    \includegraphics[width=12 cm]{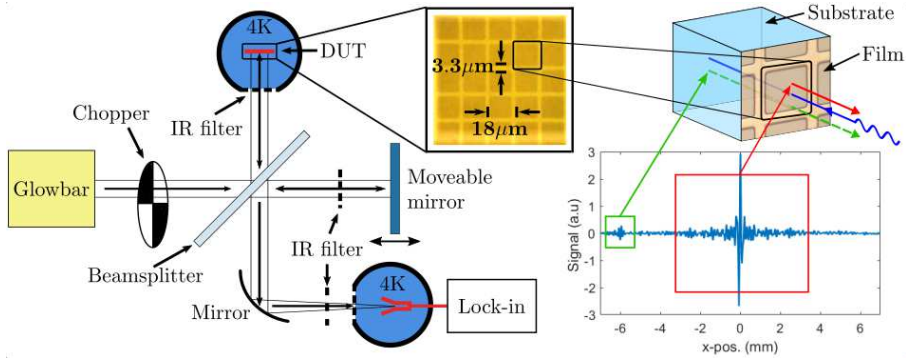}
    \caption{Schematic of the setup (left) and extraction of the film specific interferogram (right). The signal passes from a collimated glowbar through the chopper onto the beamsplitter. The signal is divided over the arms and is reflected, to interfere and be detected by a bolometer. The insert shows a picture of the meshed film, see the text for details, and a schematic of a single cell is shown. The light from the FTS (in blue) falls on the film, the majority is reflected (in red), resulting in the white light fringe. There is also reflection from the film back surface, causing a second interference peak (in green). This effect was mitigated by adding extra wafers the back surface, and the film specific interferogram was extracted (in orange).}
    \label{fig:fig1}
\end{figure*}

In this paper, we demonstrate a Terahertz DFTS scheme (Fig. \ref{fig:fig1}), which is a variation on the Fourier Transform Spectrometer (FTS) \cite{Birch1987}. The key difference is that the stationary mirror is replaced by a (dispersive) device under test (DUT). A collimated glowbar acts as a broadband radiation source. The generated signal is divided over two arms by a beamsplitter (55 $\mu$m Mylar). One part of the signal hits the movable mirror, while the other is reflected of a cryogenic cooled DUT. Here, the DUT is a superconducting film consisting of a superconductor and a substrate. Its temperature, $T$, can be increased above the critical temperature of the superconductor, $T_c$. The reflected signals of both arms interfere at the beamsplitter. The total power of the combined signal is measured by a cryogenic bolometer (NEP $=1.2 \times 10^{-13}\text{W}/\sqrt{\text{Hz}}$  \footnote{\url{http://www.infraredlaboratories.com/Bolometers.html}}). It varies as a function of a movable mirror position, resulting in an interferogram.

The presented setup allows to distinguish the reflections of different optical interfaces (see Fig. \ref{fig:fig1}). The frequency response of the film is obtained by isolating the film specific part of the interferogram and taking its Fourier transform. The effective substrate thickness was increased by adding extra wafers to its back surface (see Fig. \ref{fig:fig1}), resulting in a total thickness $>2$ mm. As a result, we removed the reflection from the back surface of the substrate, while obtaining sufficient frequency resolution in the resulting spectrum ($\Delta\nu<20$ GHz).

\begin{figure}
    \includegraphics[width=8.5 cm]{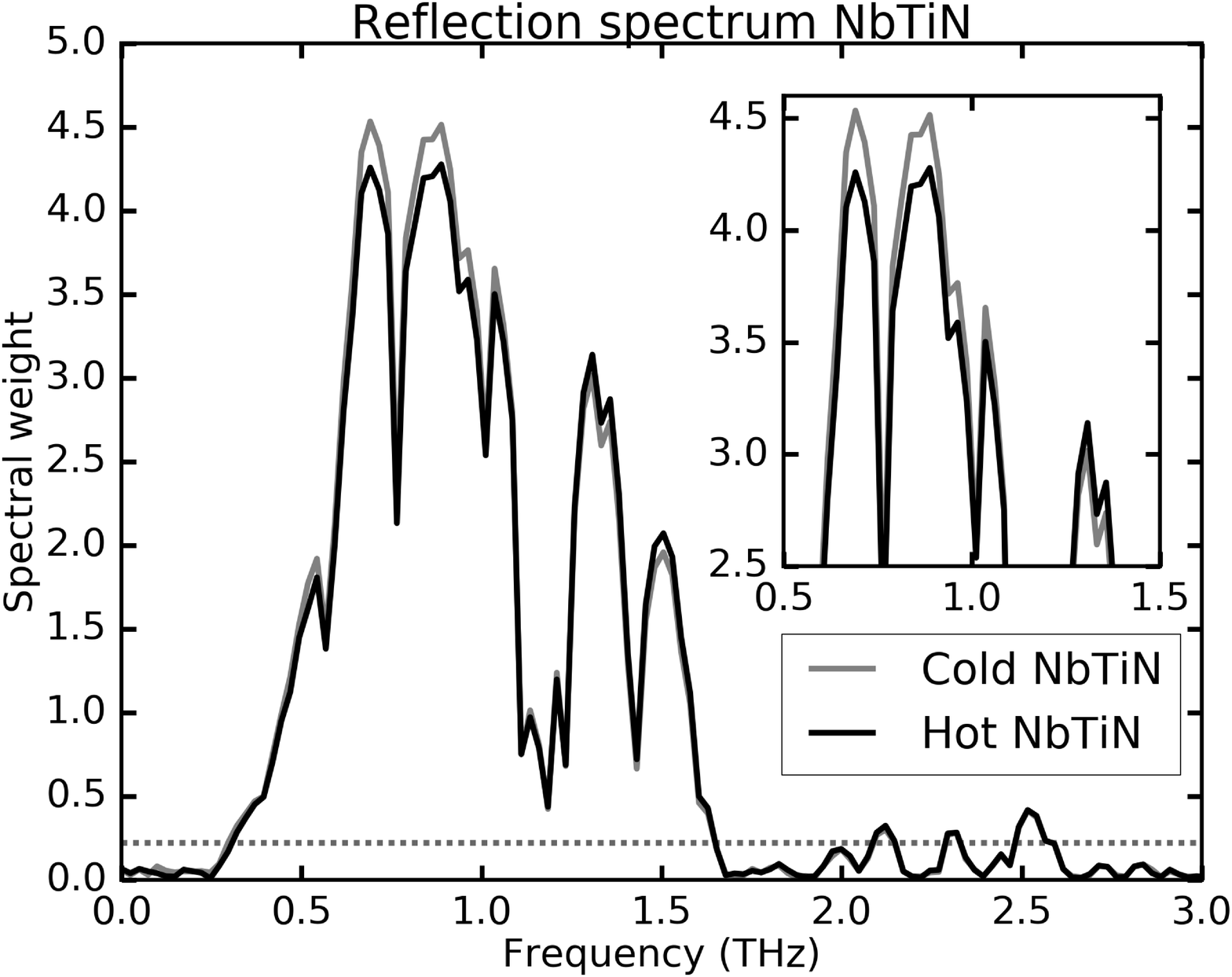}
    \caption{Cold and hot NbTiN film spectrum. The zero spectral weight around 1.9 THz is the minimum in the periodic beamsplitter function. The dotted line indicates the selected signal-to-noise threshold. The main region of interest was the gap, i.e. around 1 THz, where we have SNR of 0.2\%. For higher $\nu$, this degraded due to less signal and the stronger of water absorption lines, but this was not critical, since we were primarly interested in the gap. }
    \label{fig2}
\end{figure}

In Fig. \ref{fig2}, two NbTiN film spectra for the superconducting ($T<T_c$) and normal state ($T>T_c$) are shown. Their shapes are mainly determined by i) the detector response function, ii) infrared filters and iii) water absorption lines. These contributions are removed and the film properties are isolated, by taking the normalized difference, 
\begin{equation}
    \tilde S  = \frac{S_s - S_n }{ S_n },
\label{eq:R_contrast}
\end{equation}
where $S_s$ and $S_n$  are the power spectra taken for $T<T_c$ and $T>T_c$. In FTS theory, the intensity of the power spectrum, resulting from the interferogram, is linearly proportional to the product of field amplitudes of each arm \cite{davis2001fourier}. Thus, Eq. \ref{eq:R_contrast} describes the relative change in the square root of the film reflection coefficient, $R$, between the superconducting and the normal state in terms of power. 

\begin{table*}[b!]
\caption{Nb and NbTiN film properties. $d$ is the thickness of the film and $\sigma_{300K}$ is the DC conductivity measured at room temperature.}
\begin{ruledtabular}
\label{tab:results1}
\begin{tabular}{ ccccc }
    \# & Film    & Substrate 	& d(nm) & $\sigma_{300K}$(S/m)	\\ \hline 
    1 & Nb\footnote{Meshed film}      & Silicon & 70 	& $6.7\times10^{6}$  \\ 
    2 & Nb$^{\text{a}}$      & Silicon & 200 & $6.3\times10^{6}$ \\  
    3 \& 4 & NbTiN\footnote{Solid films}   & Silicon \& Quartz & 360 & $9.3\times10^{5}$  \\
    5 & NbTiN$^{\text{a}}$   & Silicon  & 360 & $9.3\times10^{5}$  \\
    6 & NbTiN$^{\text{a}}$   & Quartz  & 360 & $9.3\times10^{5}$ 
\end{tabular}
\end{ruledtabular}
\end{table*}

From theory, we estimated that at frequency below the gap a solid thick film ($d > \lambda_L$) resulted in $\tilde S$ to be around 0.1 and 2\% for Nb and NbTiN, respectively. To increase $\tilde S$, the Signal-to-Noise Ratio (SNR), and to study technological processes for forming superconducting lines, we studied meshed films (inset of Fig. \ref{fig:fig1}) in addition to solid ones (see Table \ref{tab:results1}). The width was $\sim 3-4\mu$m, such that it could be reliably produced by the selected technological process, and the period was $18 \mu$m, such that is $<c/10 \nu_c$, with $\nu_c =1.5$ THz being the main frequency of interest. As a result, the effective conductivity was expected to decrease by a factor $\sim 5$ for both films and $\tilde S$ increased accordingly. In case of NbTiN films, the increase in contrast was less due to higher reflection from the substrate. All of these substrate characteristics were taken into account by the simulations, as will be discussed later. 

\begin{figure}[b!]
    \includegraphics[width=8.5 cm]{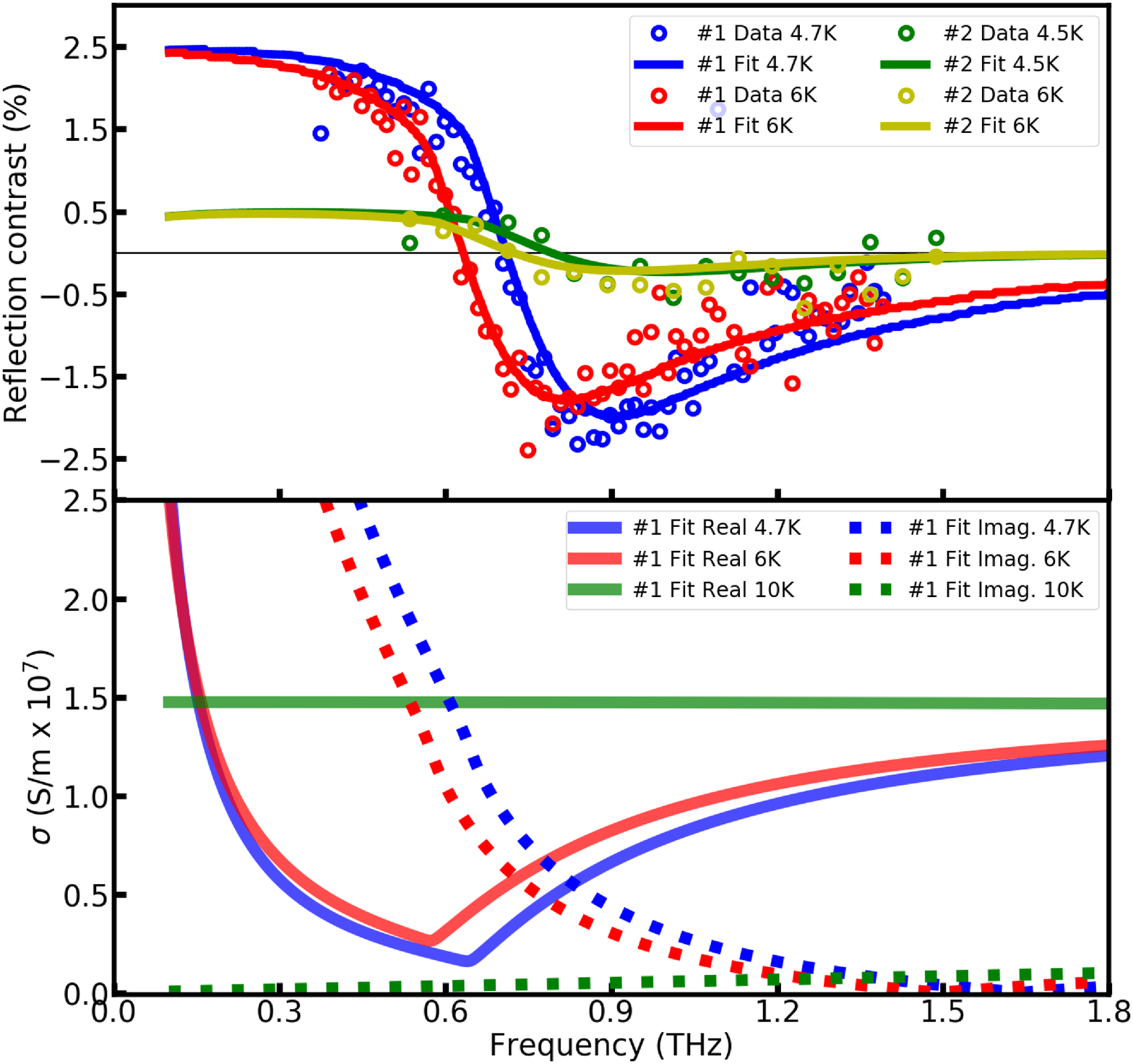}
    \caption{Nb films, i.e. films $\#1$ and $\#2$, contrast data compared with model (top) and the obtained conductivity from the fit (bottom). In the latter, the real (solid line) and the imaginary (dotted line) part of the conductivity for the superconducting and the normal state of the films are shown. Here, the curves for 6 K are omitted for clarity.}
    \label{fig3}
\end{figure}

\begin{figure}
    \includegraphics[width=8.5 cm]{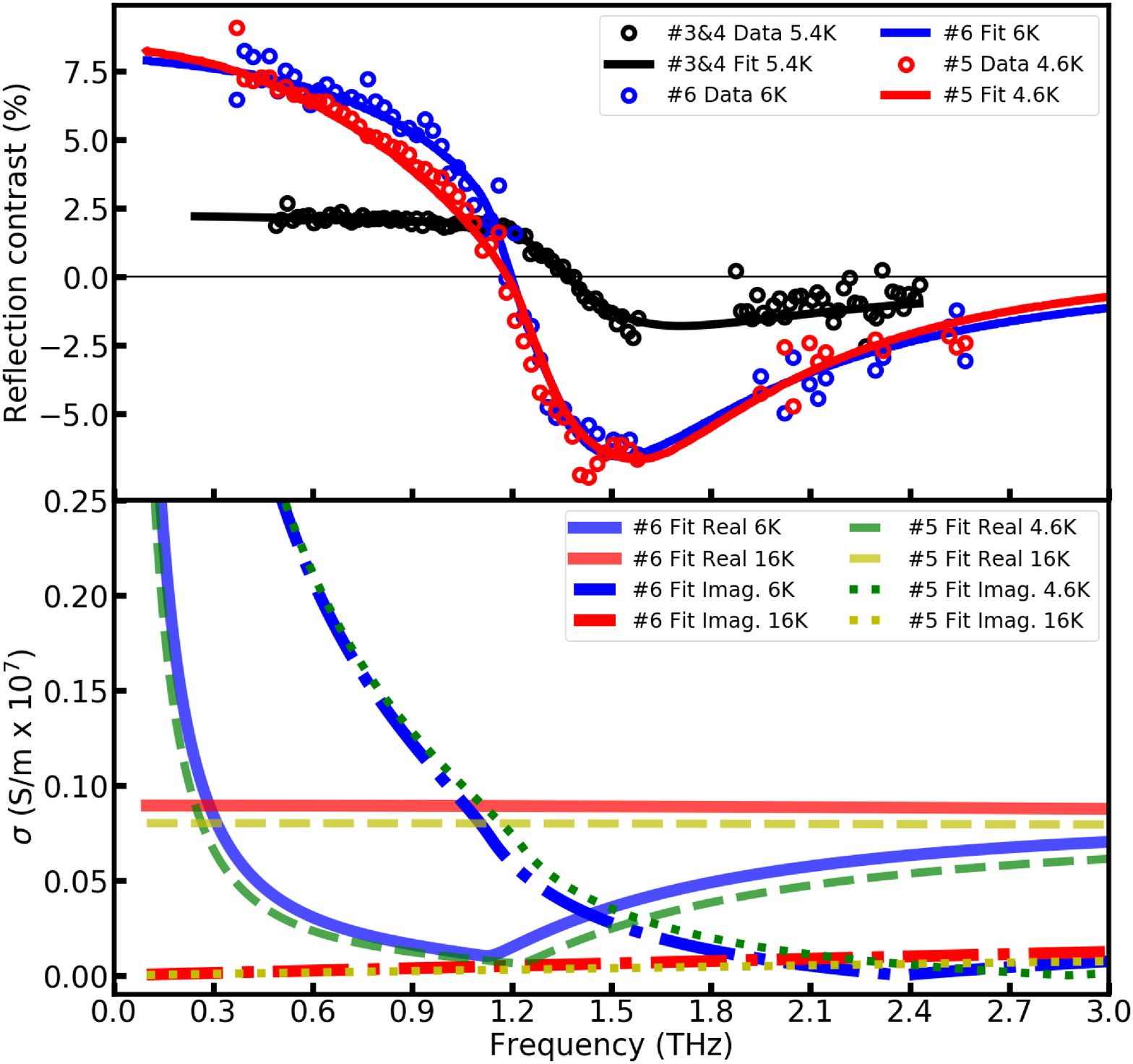} 
    \caption{Contrast data compared with model (top) and the conductivity obtained from the fit (bottom) for NbTiN 360nm film on Si and quartz substrate, i.e. films $\#3\&4$, and $\#5$ and $\#6$. In the bottom plot, the real (solid line) and the imaginary (dotted line) part of the conductivity for the superconducting and the normal state of the films are shown.}
    \label{fig4}
\end{figure}

The measured contrast curves are shown in the top row of Fig. \ref{fig3} and \ref{fig4} by the circles. At frequencies corresponding to photon energies below the superconducting gap frequency, $\nu_g$, of the cooper pairs, i.e. $\nu < \nu_g$,  the contrast is positive. Above the gap frequency ($\nu > \nu_g$) the contrast is negative, due to an increased absorption of photons and the availability of high density of states that capture unpaired electrons. Finally, far above the gap ($\nu >> \nu_g$) the reflectively for superconducting state converges to the normal state one and the contrast approaches zero.

A best fit of our model to experimental contrast curves were made and the properties of the superconducting films were extracted.  In the model, $\tilde S$ was a product of $R$, which was determined by the complex surface impedance, $Z_s$, for a particular film state. The fitting process consists of five steps.

In the first step, we calculate the normal state conductivity. The simplest model of the normal state conductivity that includes the scattering of electrons is the Drude model \cite{ASM,Lloyd-Hughes2012}, which was demonstrated to be an important parameter when describing superconducting films \cite{Uzawa2015}. The normal state AC conductivity is derived to be\cite{ASM}  
\begin{equation}
\sigma_n (\nu) = \frac{\sigma}{1 - 2 \pi i \nu \tau },
\label{eq:2}
\end{equation}
where $\nu$ is the frequency, $\sigma$ is the normal state conductivity, and $\tau$ is the electron relaxation time. 

In the second step, we calculate the ratio of the superconducting state conductivity to the normal state conductivity,
\begin{equation}
    \frac{\sigma_s}{\sigma_n} = f(\nu, d, T, \alpha, \delta, \tau),
    \label{eq:conductivity_curves}
\end{equation}
where $d$ is the film thickness and $T_c$ is the critical temperature measured at DC. Furthermore, $\delta$ is the complex gap parameter \cite{Noguchi2012}, such that 
\begin{equation}
    \Delta = \Delta_0(1 + i \delta) ,
\end{equation}
which determines the intra-gap states in strong-coupled superconductors. Mattis-Bardeen (MB) theory dictates that for a superconductor $\alpha = 2\Delta_0 / k_b T_c$, where $\alpha$ is the strong-coupling coefficient \cite{Dover82,Wang96} and $2\Delta_0$ is the energy gap. For Niobium $\alpha \approx 3.53$ \cite{Turneaure91} and even stronger coupling ($\alpha > 3.53$) was estimated for NbN \cite{Dover82,Wang96}. Equation \ref{eq:conductivity_curves} was obtained using standard MB theory with a simplifying approximation  \cite{MattisBardeen1958}, which is valid in the local limit for strongly-coupled superconductors \cite{nam1967theory} . Moreover, the model was extended by including $\delta$\, \cite{Noguchi2012}, which was required for an accurate description of the investigated superconducting films. To obtain $\sigma_s$, we multiplied the result from Eq. \ref{eq:conductivity_curves} with the normal state conductivity. 

In the third step, we calculate $Z_s$ of a solid film for both the superconducting and normal state using formula\cite{Hartemann1992,Booth1994,Noguchi2012} 
\begin{equation}
Z_s(\nu) = \sqrt{\frac{2 i \pi \nu \mu_0}{\sigma_s }} coth( \sqrt{ 2 \pi i \nu \mu_0 \sigma_s} d). 
\label{eq:zs}
\end{equation} 

This approximation is valid for both Nb and NbTiN films, because the electron mean free path is below $1$\,nm, i.e. more than 40 and 200 times, correspondingly, shorter than the skin depth. 

In the fourth step, having $Z_s$ $(= R_s + i X_s)$ and the geometry of the meshed film, the reflection coefficient of the meshed film, $R$, is determined using an electromagnetic simulator (CST Microwave Studio). The electromagnetic periodic structure of a single cell is exploited and modelled according to the Floquet theorem \cite{Floquet1883}, containing all the 3D geometric and electromagnetic properties. Then, $R= g(R_s,X_s,\nu)$ is calculated for the entire film using the Floquet boundary conditions. 

In the fifth and final step, we obtain the simulated DFTS spectra by taking the square-root of $R$ and $\tilde S$ is calculated using Eq. \ref{eq:R_contrast}, which is compared to the measurment data. 

The contrast curve for the solid film are calculated in a similar way. First, the conductivity of the superconducting film and its surface impedance are determined using Eq.\ref{eq:2}-\ref{eq:zs}. Then $R$ is estimated by 
\begin{equation}
    R= \Big| \frac{Z-Z_0}{Z+Z_0} \Big|^2,
    \label{eq:R_free_space}
\end{equation} 
where $Z_0$=377$\Omega$ is the impedance of free-space. This provides $R$ for the normal and superconducting states, and results in $\tilde S$.

Before measuring the NbTiN films, we verified the method using two Nb meshed films ($d=70$\,nm and $d=200$\,nm), since Nb is a superconductor with a well-known parameters and is easy to fabricate. The Nb films were deposited on a silicon substrate at room temperature by DC sputtering with a Nb target in an argon atmosphere. The mesh geometry (Fig. \ref{fig:fig1}) was obtained by a reactive ion etching (RIE) process. Their critical temperatures were measured to be $9 \pm 0.05$ K and $9.35 \pm 0.05$ K, respectively, using a 4-point DC connection. A $T_c = 9.0$ K is considered to be normal for thin Nb films \cite{Gubin2005}. The DC measurements also provided the room temperature conductivities, $\sigma_{300K}$, and the RRR levels (see Table \ref{tab:results2}).

\begin{table*}
\caption{Nb and NbTiN film parameters determined by the DC measurement and RF fit. $T_c$, $\sigma_{dc,cold}$ and $\text{RRR}$ are the critical temperature, the normal state conductivity and the ratio $\sigma_{dc,cold}/\sigma_{300K}$ determined from DC measurements. $\sigma$, $\alpha$, $\delta$ and $\tau$ are obtained from the RF fit, and are the conductivity just above the critical temperature, the strong-coupling coefficient, the complex gap parameter and the Drude relaxation time.}
\begin{ruledtabular}
\label{tab:results2}
\begin{tabular}{ cccc|cccc }
    \multicolumn{4}{c|}{\textbf{DC measurement}} & \multicolumn{4}{c}{\textbf{RF fit}}  \\ 
    \# & $T_c$& $\sigma_{dc,cold}$(S/m) & $\text{RRR}$ & $\sigma$ ($\times10^{5}$ S/m) & $\alpha$ & $\delta$	& $\tau$ (fs)	\\ \hline 
    1 & $9.0 \pm 0.05$ & $20.8\times10^{6}$ & 3.1 & $150 \pm 14$ & $3.52 \pm 0.02$ & $5.0\pm 1.7\times10^{-3}$ & $4 \pm 5$\\ 
    2 & $9.35\pm 0.05$ & $25.0\times10^{6}$ & 4.0 & $150 \pm 14$ & $3.52 \pm 0.02$ & $5.0 \pm 1.7\times10^{-3}$ & $4 \pm 5$\\  
    3\&4 & $ 14.7 \pm 0.1$ & $9.1\times10^{5}$ & 0.98 & $6.5 \pm 0.7$ & $3.86 \pm 0.04$ & $2.0 \pm 1.5\times10^{-2}$ & $10 \pm 5$ \\
    5 & $14.65 \pm 0.1$ & $9.1\times10^{5}$ & 1.01 & $8.0 \pm 0.7 $ & $4.02\pm 0.04$ & $1.0 \pm 1.5 \times10^{-2}$ & $5 \pm 5 $\\
    6 & $14.6 \pm 0.1$ & $9.4\times10^{5}$ & 1.01 & $8.9 \pm 0.7$ & $3.71\pm 0.04$ & $1.5 \pm 1.5 \times10^{-2}$ & $7.5 \pm 5$
\end{tabular}
\end{ruledtabular}
\end{table*}

The measured contrast curves for the Nb films at different temperatures are shown in the top plot of Fig. \ref{fig3}, where the experimental data is indicated by the circles. Some of the frequency data points were removed, since they did not meet the SNR threshold (see Fig. \ref{fig2}). The contrast for $70$\,nm film is much higher than for $200$\,nm, due to the higher resistance of the thin film in the normal state.

In this figure, the best fit of the model to the Nb films contrast curves are indicated by the solid curves. The fit was done by minimizing the root-mean-square (RMS) difference between the measured and simulated contrast curve. Four parameters were used for the fit: $\sigma$, $\alpha$, $\delta$ and $\tau$. Here, $\sigma$ scales the contract curve along the vertical axis and is set by fitting the contrast-level to the experimental data at low $\nu$; $\delta$ is responsible for the steepness of the curve near the gap, where the contrast crosses zero; $\tau$ mainly affects the shape of the curve above the gap and its depth right after the gap (i.e. the so-called "dip" around 0.9 THz for Nb and 1.5 THz for NbTiN), and $\alpha$ scales the contrast curve along the horizontal axis, because it controls the energy of superconducting gap, which is obtained by $\alpha T_c$. 

The minimization procedure comprised three steps. In the first step, we made an initial guess of the parameters using the DC measurements ($\sigma$ and $\alpha$) or the values we found in the literature ($\delta$ and $\tau$). In the second step, a Monte Carlo simulation was carried out for each parameter, and the normalized (convex) RMS curve was determined, which gave the local minimum parameter value. This procedure was repeated iteratively until the global minimum was found, and the final RF fit parameters were obtained. In the third and final step, a Monte Carlo simulation was carried out using the obtained RF fit parameters, to estimate the uncertainties in the parameters, which was the parameter value that correspond with a RMS deviation of 10\%. The obtained RF fit parameters and their errors are shown in Table \ref{tab:results2}, and the real and imaginary part of the obtained conductivity curves are given in the bottom plot of Fig. \ref{fig3} for completeness.

The $\sigma$ obtained from fitting for both the $70$\,nm and $200$\,nm Nb film were $15$\,S/m, which is significantly lower than for the DC measurements. In the latter, we probe $\sigma_{dc,cold}$ through the lowest resistance path, which is only a localized part of the film. On the other hand, in the RF test we are sensitive to the averaged $\sigma$ of the entire film. Probably, this explains the difference between $\sigma_{dc,cold}$ and $\sigma$. This is an important result and should be taken into account when designing future RF Nb mixer based circuits. Furthermore, their respective values for $\alpha$ agreed very well with $\alpha=3.53$, and there was a fairly good agreement between the literature values for $\delta$ \cite{noguchi2010, Noguchi2012} and $\tau$ \cite{Yoo1990,Romaniello2006}. Based on these results we concluded that our DFTS characterisation scheme worked for Nb films, and that we could apply it to NbTiN films.

The NbTiN films were deposited on substrates of two different materials: silicon (Si) and fused quartz (Qu), and for both a solid and meshed film were made. See Table \ref{tab:results1} for the labeling of these films. Deposition was done at room temperature by DC sputtering using a NbTiN target in a mixture of nitrogen and argon atmosphere \cite{Khudchenko2016}. Film $\#5$ and $\#6$ were deposited together in one sputtering cycle and for both $d = 360$ nm. Films $\#3\&4$, i.e. the solid films, were deposited later, but using a similar procedure. The geometry of the mesh was the same as for the Nb films and the same RIE process was used. The DC tests results are shown in Table \ref{tab:results2} and the RRR level was found to be very close to 1, as expected.

The reflection measurement results for the NbTiN films are shown in Fig. \ref{fig4}. The contrast curves for $\#3\&4$ were very similar, hence, we only show one of them here to avoid confusion. From this figure, we see that the contrast level for NbTiN films is much higher than for Nb films, due to lower conductivity. 

The fitting of the model to the NbTiN contrast curves (Fig. \ref{fig4}) was done in the same way as for the Nb films. The obtained fitting parameters are listed and compared with the DC measurements in Table \ref{tab:results2}. Here, the results for film $\#3\&4$ are combined, because they were very similar. The obtained values for $\tau$ are in good correspondence with the value of $10$ fs given in literature \cite{Gousev1994}. We also found that $\delta$ agreed well with the literature value\cite{Noguchi2012}. Finally, the obtained $\alpha$'s for all NbTiN films, i.e. films  $\#3\&4$, $\#5$, and $\#6$, show a deviation from $3.53$, the typical value for Nb. Still, all the fitting values fall well within the ranges mentioned in the literature. For instance, for NbN $\alpha$ was shown to be even $4.25$ \cite{Dover82}. For completeness, we included the real and imaginary part of the fitted conductivity curves for films $\#5$ and $\#6$ in Fig. \ref{fig4}.

The DC measurements suggest that the quality of the meshed films (i.e. $\#5$ and $\#6$) is higher than for the solid films (film $\#3\&4$) due to a higher $T_c$. This is inline with the RF measurements demonstrating a higher superconducting energy gap due to a higher $\alpha$ (see Table \ref{tab:results2}). This is an important outcome, because it means that we see no deterioration of the NbTiN film due to the formation processes, and the presented films can be used as a bottom electrode in SIS mixers up to a frequency 1\,THz.
In addition, the meshed film deposited on the silicon substrate (film $\#5$) has a higher $\alpha$ value than the one deposited on quartz (film $\#6$), while the $T_c$ values obtained by the DC measurements are almost identical. The observed higher superconducting gap of film $\#5$ is explained by a better match between the crystal lattice structure of the NbTiN film and that of the silicon substrate opposed to the lattice match for the quartz substrate. This result demonstrates that presented DFTS scheme is indeed a sensitive and effective tool for determining the characteristic parameters of superconducting films, and that it can become an important quality-control tool for superconducting structures. 

To conclude, we have developed a Dispersive Fourier Transform Spectrometer scheme and used it for the characterization of superconducting films by directly determining their key parameters, such as the complex conductivity and the strong-coupling coefficient, at THz frequencies. In future work, we will apply this method to the thick films used for the fabrication of THz mixers, and we plan to investigate their properties further by analyzing their dispersive properties using the Kramers-Kroning relations.

This work was supported by Netherlands Academy of science (NWO) grant “NWO-FAPESP Advanced Instrumentation for Astronomy” (Project No. 629.004.00). The development of the NbTiN films was supported by the Russian Science Foundation (Project No. 19-19-00618). The numerical simulations were supported by the Russian Science Foundation (Project No. 17-79-20343). The equipment of USU "Cryointegral" was used to carry out the research; USU is supported by a grant from the Ministry of Science and Higher Education of the Russian Federation, agreement No. 075-15-2021-667.

\section*{Data Availability}
The data that support the findings of this study are available from the corresponding authors upon reasonable request.

\nocite{*}
\bibliography{main}
\end{document}